# Drop-port study of microresonator frequency combs: power transfer and time-domain characterization

Pei-Hsun Wang[1,*], Yi Xuan[1,2], Li Fan[1,2], Leo Tom Varghese[1,2], Jian Wang[1,2], Yang Liu[1,], Daniel E. Leaird[1], Minghao Qi[1,2], and Andrew M. Weiner[1,2]

[1]*School of Electrical and Computer Engineering, Purdue University,465 Northwestern Avenue, West Lafayette, IN 47907-2035, USA*
[2]*Birck Nanotechnology Center, Purdue University, 1205 West State Street, West Lafayette, Indiana 47907, USA*
*\* wang1173@purdue.edu*

**Abstract:** We use a drop-port geometry to study power transfer in silicon nitride on-chip microresonator frequency comb generators. In sharp contrast with the traditional transmission geometry, we observe smooth output spectra with comparable powers in the pump and adjacent comb lines. An observation of saturation in the drop-port output power is explained semi-empirically by introducing pump saturation into a coupling of modes model. Autocorrelation measurements are performed on the drop-port output, without the need to filter out or suppress the strong pump line as is necessary in thru-port experiments. Passively mode-locked pulses are observed with a normal dispersion microcavity.

**References and links**

1. P. Del'Haye, A. Schliesser, O. Arcizet, T. Wilken, R. Holzwarth, and T. J. Kippenberg, "Optical frequency comb generation from a monolithic microresonator," Nature **450**, 1214-1217 (2007).
2. I. S. Grudinin, N. Yu, and L. Maleki, "Generation of optical frequency combs with a CaF$_2$ resonator," Opt. Lett. **34**, 878-880 (2009).
3. D. V. Strekalov and N. Yu, "Generation of optical combs in a whispering gallery mode resonator from a bichromatic pump," Phys. Rev. A **79**, 041805 (2009).
4. J. S. Levy, A. Gondarenko, M. A. Foster, A. C. Turner-Foster, A. L. Gaeta, and M. Lipson, "CMOS-compatible multiple-wavelength oscillator for on-chip optical interconnects," Nat. Photonics **4**, 37-40 (2010).
5. L. Razzari, D. Duchesne, M. Ferrera, R. Morandotti, S. Chu, B. E. Little, and D. J. Moss, "CMOS-compatible integrated optical hyper-parametric oscillator," Nat. Photonics **4**, 41-45 (2010).
6. Y. Chembo and N. Yu, "Modal expansion approach to optical-frequency-comb generation with monolithic whispering-gallery-mode resonators," Phys. Rev. A, **82**, 033801 (2010).
7. W. Liang, A. A. Savchenkov, A. B. Matsko, V. S. Ilchenko, D. Seidel, and L. Maleki, "Generation of near-infrared frequency combs from a MgF$_2$ whispering gallery mode resonator," Opt. Lett. **36**, 2290-2292 (2011).
8. T. J. Kippenberg, R. Holzwarth, and S. A. Diddams, "Microresonator-based optical frequency combs," Science **332**, 555–559 (2011).
9. F. Ferdous, H. Miao, D. E. Leaird, K. Srinivasan, J. Wang, L. Chen, L. T. Varghese, and A. M. Weiner, "Spectral line-by-line pulse shaping of on-chip microresonator frequency combs," Nat. Photonics **5**, 770-776 (2011).
10. S. B. Papp and S. A. Diddams, "Spectral and temporal characterization of a fused-quartz-microresonator optical frequency comb," Phy. Rev. A **84**, 053833 (2011).
11. I. S. Grudinin, L. Baumgartel, and N. Yu, "Frequency comb from a microresonator with engineered spectrum," Opt. Express **20**, 6604-6609 (2012).
12. T. Herr, K. Hartinger, J. Riemensberger, C. Y. Wang, E. Gavartin, R. Holzwarth, M. L. Gorodetsky, and T. J. Kippenberg, "Universal formation dynamics and noise of Kerr-frequency combs in microresonators," Nat. Photonics **6**, 480-487 (2012).
13. F. Ferdous, H. Miao, P.-H. Wang, D. E. Leaird, K. Srinivasan, L. Chen, V. Aksyuk, and A. M. Weiner, " Probing coherence in microcavity frequency combs via optical pulse shaping," Opt. Express **20**, 21033–21043 (2012).

14. T. Herr, V. Brasch, J. D. Jost, C. Y. Wang, N. M. Kondratiev, M. L. Gorodetsky, and T. J. Kippenberg, "Temporal solitons in optical microresonators," arXiv : 1211.0733v3 (2013).
15. K. Saha, Y. Okawachi, B. Shim, J. Levy, R. Salem, A. Johnson, M. Foster, M. Lamont, M. Lipson, and A. Gaeta, "Modelocking and femtosecond pulse generation in chip-based frequency combs," Opt. Express **21**, 1335-1343 (2013).
16. W. Liang, V. S. Ilchenko, A. A. Savchenkov, A. B. Matsko, D. Seidel, and L. Maleki, "Passively mode locked Raman laser," Phys. Rev. Lett. **105**, 143903 (2010).
17. T. Barwicz, M. Popović, P. Rakich, M. Watts, H. Haus, E. Ippen, and H. Smith, "Microring-resonator-based add-drop filters in SiN: fabrication and analysis," Opt. Express **12**,1437-1442 (2004).
18. T. Barwicz, M. Popović, M. Watts, P. Rakich, E. Ippen, and H. Smith, "Fabrication of add-drop filters based on frequency-matched microring resonators," J. Lightwave Technol. **24**, 2207-2218 (2006).
19. M. Peccianti, A. Pasquazi, Y. Park, B. E. Little, S. T. Chu, D. J. Moss, and R. Morandotti, "Demonstration of a stable ultrafast laser based on a nonlinear microcavity," Nat. Commun. **3**, 765 (2012).
20. A. Pasquazi, M. Peccianti, B. Little, S. Chu, D. Moss, and R. Morandotti, "Stable, dual mode, high repetition rate mode-locked laser based on a microring resonator," Opt. Express **20**, 27355-27363 (2012).
21. Z. Jiang, C. Huang, D. E. Leaird, and A. M. Weiner, "Optical arbitrary waveform processing of more than 100 spectral comb lines," Nat. Photonics **1**, 463-467 (2007).
22. S. T. Cundiff and A. M. Weiner, "Optical arbitrary waveform generation," Nat. Photonics **4**, 760-766 (2010).
23. J. Čtyroký, I. Richter, and V. Šinor, "Dual resonance in a waveguide-coupled ring microresonator," Optic. Quantum Electron. **38**(9–11), 781–797 (2010).
24. B. Little, J. Laine, and S. Chu, "Surface-roughness-induced contradirectional coupling in ring and disk resonators," Opt. Lett. **22**, 4-6 (1997).
25. S. Johnson and J. Joannopoulos, "Block-iterative frequency-domain methods for Maxwell's equations in a planewave basis," Opt. Express **8**, 173-190 (2001).
26. T. Carmon, L. Yang, and K. Vahala, "Dynamical thermal behavior and thermal self-stability of microcavities," Opt. Express **12**, 4742-4750 (2004).
27. A. Matsko, A. Savchenkov, W. Liang, V. Ilchenko, D. Seidel, and L. Maleki, "Mode-locked Kerr frequency combs," Opt. Lett. **36**, 2845-2847 (2011).
28. S. Xiao, M. Khan, H. Shen, and M. Qi, "Modeling and measurement of losses in silicon-on-insulator resonators and bends," Opt. Express **15**, 10553-10561 (2007).
29. M. Ebrahimzadeh, G. Turnbull, T. Edwards, D. Stothard, I. Lindsay, and M. Dunn, "Intracavity continuous-wave singly resonant optical parametric oscillators," J. Opt. Soc. Am. B **16**, 1499-1511 (1999).
30. F. Colville, M. Dunn, and M. Ebrahimzadeh, "Continuous-wave, singly resonant, intracavity parametric oscillator," Opt. Lett. **22**, 75-77 (1997).
31. A. Matsko, A. Savchenkov, and L. Maleki, "Normal group-velocity dispersion Kerr frequency comb," Opt. Lett. **37**, 43-45 (2012).
32. A. Savchenkov, A. Matsko, W. Liang, V. Ilchenko, D. Seidel, and L. Maleki, "Kerr frequency comb generation in overmoded resonators," Opt. Express **20**, 27290-27298 (2012).

## 1. Introduction

Optical frequency combs based on nonlinear optical modulation in high quality factor (Q) microresonators, such as toroidal microresonators, crystalline resonators, and waveguide ring resonators, are the subject of considerable attention [1-15]. Through cascaded four wave mixing (FWM) in these high Q devices, the built-up intracavity power enables additional cavity modes to oscillate. In most previous studies a thru-port geometry consisting of a single bus waveguide or fiber is used to couple light both into and out from the microresonators, while a few experiments utilize a drop-port as an output coupler [2, 3, 5, 11, 16]. Typically, the comb spectrum observed at the thru-port contains a strong peak at the pump frequency, which often is 20 dB or more above the adjacent comb lines [1, 4, 7-10, 12-15]. Thus, the power transfer into the comb is often poor. To shed light on these issues, here we study comb generation in a silicon nitride microresonator fabricated with a drop-port. In addition to investigations of microresonator-based combs, a drop-port geometry has been widely used in optical add-drop filters [17, 18] and has also been employed in microresonator-assisted mode-locked lasers [19, 20]. By observing the output from the drop-port, we can directly probe the comb fields internal to the microring. This avoids a complication inherent at the thru-port, for which the output pump light depends on the interference between the field coupled out of the microresonator and the field directly transmitted from the input bus. Although the introduction of a drop-port does impact the loaded quality factor, it also provides new information.

Our work results in several new findings. First, the comb spectra observed at the drop-port are smooth, with the intensity of the pump line comparable to that of the comb lines. Smooth spectra such as reported here are potentially advantageous for line-by-line pulse shaping studies [21, 22]. Second, the (weak) power transfer into the comb may be explained to a large degree by the coupling parameters characterizing the linear operation of the resonances studied. Third, accounting for the readily observed pump saturation in a coupling of modes model modifies the pump power coupled into the resonator and improves the fit with the drop-port output power data, for which saturation behavior are also observed. These findings provide insight relevant to design the optimum coupling in the presence of comb generation. Finally, by performing autocorrelation measurements to characterize the time-domain behavior of the combs from the drop-port, we observe direct ultrashort pulse generation, without the need to filter out or suppress the strong pump as was necessary in previous thru-port experiments [9, 10, 13-15]. Although several recent studies have reported ultrashort pulse generation with microresonator-based combs, either directly [14, 15] or with phase compensation via line-by-line pulse shaping [9, 10, 13], this paper is the first, to the best of our knowledge, to report ultrashort pulse generation directly from a microcavity in the normal dispersion regime.

The remainder of this paper is organized as follows. Section 2 gives a brief overview of the experimental setup. Section 3 compares comb spectra at thru- and drop-ports and investigates power transfer. Time-domain measurements are presented in Section 4. In Section 5, we conclude.

## 2. Experimental setup

Figure 1(a) shows a microscope image of the silicon nitride microring resonator with 2 µm × 550 nm waveguide cross-section and 100 µm radius. The simulated dispersion parameter, including both waveguide and material dispersion, is -306 ps/nm/km, which is in the normal dispersion regime. The gaps between the resonator and the thru-port and drop-port coupling waveguides are both 700 nm. For stable coupling on- and off-chip, lensed single mode fibers are placed into U-grooves as shown in Fig. 1(b) and aligned with the waveguide. We note that although in the current experiments, the thru- and drop-port coupling are symmetric, asymmetric designs with weaker drop-port coupling are also possible. This would lessen the effect on the loaded Q, at the cost of reduced power transfer to the drop-port, which is appropriate if the primary aim is to provide a monitor for the internal comb field.

Figure 1(c) shows the block diagram measurement setup. Measurements of spectra at thru- and drop-ports are performed sequentially using an optical spectrum analyzer (OSA); the lensed fiber is positioned first at one output and then moved to the other, typically with the input lensed fiber fixed. Time-domain measurements were performed using intensity autocorrelation based on second harmonic generation in a noncollinear geometry. For time-domain characterization, the output from the drop-port is sent into an erbium-doped fiber amplifier (EDFA). The amplified output is then sent through a length of dispersion compensating fiber (DCF) and connected to the intensity autocorrelator. The length of the DCF was set by injecting a short test pulse laser from a passively mode-locked fiber laser through the fiber link (starting at the output lensed fiber). The DCF was trimmed to minimize the duration of the autocorrelation trace. The autocorrelation width of the test pulse after the fiber link was 368 fs full-width at half-maximum (FWHM). This is close to the autocorrelation width (302 fs) calculated using the optical spectrum after the EDFA with the assumption of flat spectral phase. The slight difference likely results from residual higher-order dispersion. We conclude that for pulse durations down to several hundred femtoseconds, the link is approximately dispersion compensated, so that autocorrelation measurements reflect the actual field at the microresonator.

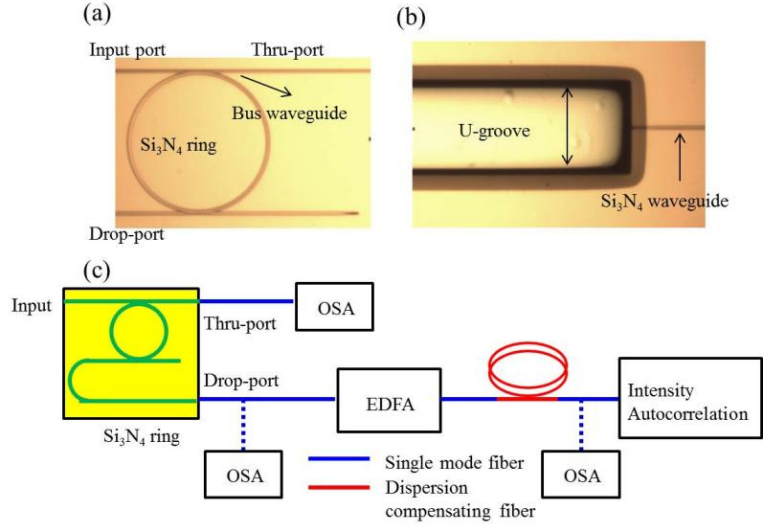

Fig. 1. (a) Microscope image of a 100 μm radius silicon nitride microring with coupling region. (b) Image of a U-groove. (c) Experimental setup for time-domain characterization.

## 3. Comb spectra and power transfer

A tunable continuous-wave (CW) laser is amplified and launched into the ring resonator. For our study of comb spectra and power transfer, we investigate two TM resonances, for which the electric field is predominantly polarized perpendicular to the plane of the wafer. The first resonance ($TM_1$) is a doublet centered at ≈1558.4 nm (Fig. 2(a) inset), has intrinsic quality factor (Q) approximately $3.1 \times 10^6$ for each of the individual lines, and belongs to a mode family with 1.84 nm average free spectral range (FSR). This splitting of the resonant peak may result from a backward propagating wave induced by scattering from the coupling regions [23] or surface roughness [24]. The second resonance (Fig. 2(b) inset) occurs near 1558.7 nm, has intrinsic Q around $1.0 \times 10^6$, and belongs to mode $TM_2$ with 1.76 nm average FSR. The modes here are identified by simulating the group index for different modes [25] and comparing the estimated FSR with the experimental one.

To generate a frequency comb, the laser is tuned to a resonance from the short wavelength side to achieve a "soft thermal lock" [4, 5, 26]. Figure 2 shows the optical spectra observed both from the thru- (blue trace) and drop-port (red trace) at the highest input powers studied: (a) 430 mW at around 1558.5 nm ($TM_1$ mode) and (b) 680 mW at around 1558.8 nm ($TM_2$ mode). Here the input power is defined as the power in the input waveguide before the microring. We estimated the power in the waveguide by measuring the input- to thru-port fiber to fiber loss, typically 3~5 dB, and assigning half of this loss to the input side. For the resonances studied here, frequency combs are observed to generate lines spaced by 3 FSRs for the $TM_1$ mode and 5 FSRs for the $TM_2$ mode. Both resonances are red-shifted around 0.1 nm due to the thermal nonlinearity under the pumping conditions quoted above.

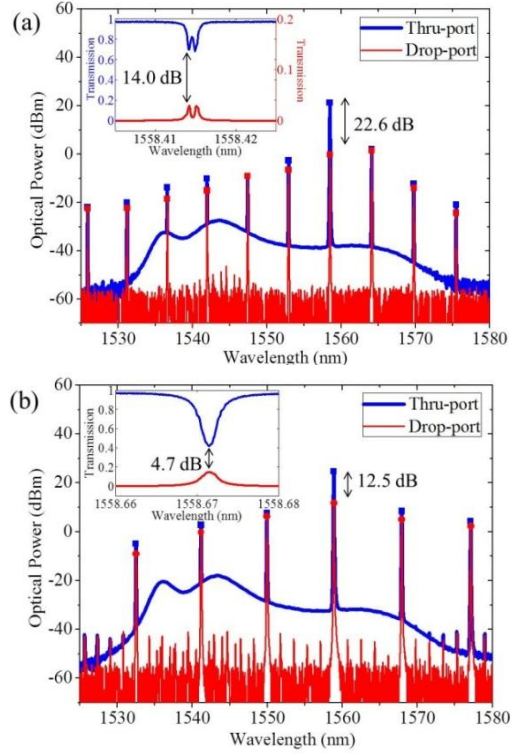

Fig. 2. Measured optical spectra of generated combs from both the thru- (blue trace) and drop-port (red trace) with input power (a) 430 mW at around 1558.5 nm ($TM_1$ mode) and (b) 680 mW at around 1558.8 nm ($TM_2$ mode). The insets show the corresponding transmission spectra with low input power.

For the spectra from the thru-port, amplified spontaneous emission (ASE) noise is clearly observed. For spectra measured from the drop-port, the ASE noise between the comb lines is filtered out. An important point is that the drop-port data reveal the power spectrum internal to the microring. Although the power at the pump frequency is much stronger than that of the adjacent comb lines in the thru-port data, in the drop-port data the power at the pump frequency is much closer to that of adjacent lines. Evidently, for the resonances studied, the combs internal to the microring have approximately smooth spectra. For the data of Fig. 2(a), the pump power at the drop-port is 22.6 dB lower than at the thru-port. For the thru-port, the pump is 19.1 dB stronger than the stronger of the adjacent lines, while at the drop-port the pump is actually slightly (1.7 dB) weaker than the stronger adjacent line. For the data of Fig. 2(b), the pump at the drop-port is 12.5 dB weaker than at the thru-port. Compared to the stronger of the adjacent lines, the pump is 16.3 dB stronger at the thru-port but only 5.5 dB stronger at the drop-port. We may conclude that to a large extent, the much stronger pump line observed at the thru-port is related to coupling conditions, not to the actual comb in the ring. Even compared to previous studies employing a drop-port geometry [2, 5, 11], our spectra, particularly Fig. 2(a) – and another example, Fig. 6(a), presented later – exhibit a degree of smoothness in the region around the pump rarely observed. An analytical investigation of mode-locked Kerr combs, predicated on anomalous group velocity dispersion, predicts a large CW background at the thru-port associated with a strong residual pump [27]. In addition, the solutions presented explicitly include a continuous-wave component internal to the microresonator. The absence of a strong pump line in our drop-port spectra suggests that at least under some conditions, possibly associated with the normal group velocity dispersion of our devices, the continuous-wave component of the internal field can be small.

Note that aside from the pump line, the output comb lines from the thru- and drop-ports in Fig. 2 exhibit close to the same power levels, except the drop-port lines are reduced by an average of 2.5 dB relative to the thru-port. Since with symmetrical coupling, the comb lines from thru- and drop-ports should have the same power, this implies either the coupling is unintentionally slightly different or the output drop guide has larger loss. To account for this difference, we adjust the estimated drop-port powers upwards by 2.5 dB in the output vs. input power data presented later as Fig. 3.

Since the large fraction of pump power that emerges from the thru-port is attributed to the coupling conditions, it is interesting to consider the coupling effects quantitatively. Following the theory proposed in Ref. [28], the coupling parameters could be extracted by fitting the thru-port transmission spectra which can be written as:

$$T_{thru} = \frac{(\lambda - \lambda_0)^2 + (\frac{FSR}{4\pi})^2 (\kappa_d^2 + \kappa_p^2 - \kappa_e^2)^2}{(\lambda - \lambda_0)^2 + (\frac{FSR}{4\pi})^2 (\kappa_d^2 + \kappa_p^2 + \kappa_e^2)^2} \quad (1)$$

where $T_{thru}$ is the power transmission response at the thru-port, and FSR is the free spectral range in the resonators in wavelength units. $\kappa_e^2$ and $\kappa_d^2$ are dimensionless coefficients characterizing the power coupled into and out of the microring from the input and drop waveguide per round trip, respectively, while $\kappa_p^2$ is the dimensionless power loss per round trip due to intrinsic loss in the resonator. The fitting results for the two resonances investigated are shown in Table 1. Note that for the $TM_1$ resonance, the $\kappa$ coefficients are calculated using the linewidth of a single resonance (short wavelength side) making up the doublet. For both resonances symmetrical coupling is assumed. The power transmission response of the drop-port $T_{drop}$ may be calculated based on the parameters extracted from Eq. (1) and is written as:

$$T_{drop} = \frac{4 \times (\frac{FSR}{4\pi})^2 (\kappa_d^2 \times \kappa_e^2)}{(\lambda - \lambda_0)^2 + (\frac{FSR}{4\pi})^2 (\kappa_d^2 + \kappa_p^2 + \kappa_e^2)^2} \quad (2)$$

**Table 1. Coupling and Transmission Parameters**

| Resonance | Calculated by Coupling Parameters | | | | Transmission Spectra | | | |
|---|---|---|---|---|---|---|---|---|
| | $\kappa_e^2$ | $\kappa_d^2$ | $\kappa_p^2$ | $T_{thru}/T_{drop}$ | $T_{thru}$ | $T_{drop}$ | $T_{thru}/T_{drop}$ (dB) | Linewidth (pm) |
| 1558.4 nm ($TM_1$) | 1.49×10⁻⁴ | 1.49×10⁻⁴ | 1.57×10⁻³ | 14.5 | 0.70 | 0.028 | 14.0 | 0.7 |
| 1558.7 nm ($TM_2$) | 1.41×10⁻³ | 1.41×10⁻³ | 5.04×10⁻³ | 5 | 0.42 | 0.142 | 4.7 | 2.3 |

As shown in Table 1, the $T_{thru}/T_{drop}$ values from Eq. (1) and (2) are in close agreement with those observed experimentally, giving confidence in the parameters extracted. We note that for both resonances, the loss in the microring is significantly larger than the waveguide couplings. Hence, the resonances are under-coupled. In this under-coupled regime, the introduction of a drop-port has relatively weak impact on the loaded Q. We also note that the intrinsic loss for the 1558.7 nm ($TM_2$) resonance is approximately three times higher than that for the 1558.4 nm ($TM_1$) resonance, consistent with the fact that the fundamental mode usually has lowest intrinsic loss. Finally, from the parameters shown in the Table, we observe that the under-coupling is much more severe for the $TM_1$ resonance.

We have studied the power distribution in the combs observed at thru- and drop-ports with varying input power. To account for thermo-optically induced resonance shifts, at each input power the pump wavelength is tuned carefully into resonance, which we have defined as the wavelength yielding minimum power transmission from the thru-port (maximum transmission from the drop-port). Figure 3(a) shows the output pump power (black trace) and comb power (blue trace) at the thru-port versus input power for the $TM_1$ resonance. The input and output powers refer to on-chip powers estimated before and after the microring, respectively. The comb power is obtained by integrating the measured optical spectrum excluding the pump line, while the pump power is determined by integrating over a 1 nm range centered on the pump. Both pump and comb power increase with input power, with the comb showing a threshold effect. The corresponding drop-port results are shown in Fig. 3(b). Unlike the nearly linear pump power curve measured at the thru-port, at the drop-port the pump power clearly saturates (although not abruptly) above the threshold for comb generation.

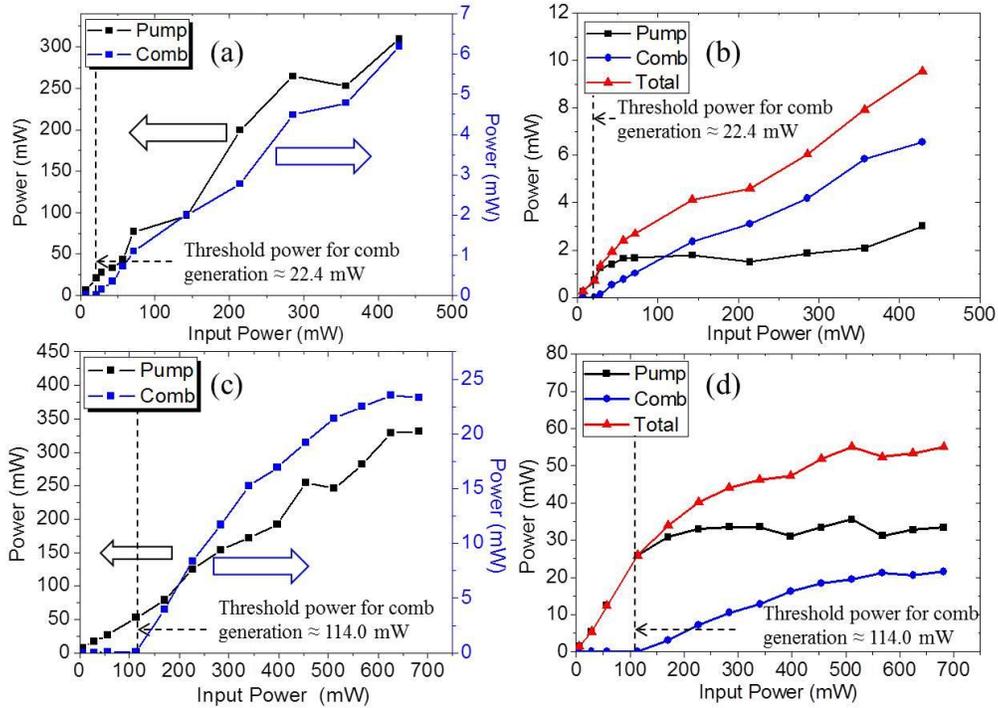

Fig. 3. The detected output power of the pump (black trace) and the comb lines other than the pump (blue trace) versus input power, corresponding to the (a) thru- and (b) drop-port at 1558.4 nm ($TM_1$) resonance while (c) and (d) represent the analogous results from the thru- and drop-port at 1558.7 nm ($TM_2$) resonance. The red traces in (b) and (d) stand for the total power including both pump and comb lines.

The $TM_2$ resonance furnishes another example of pump saturation. Figures 3(c) and 3(d) show the detected output power at the thru- and drop-port versus the input power. Clamping of the pump power in the microring above the comb generation threshold is again observed. Any further increase in the input power beyond threshold leads to an approximately linear rise – at least initially – in the parametric signals (i.e., the frequency combs). This observation suggests saturation of the parametric gain analogous to gain saturation observed in conventional parametric oscillators above threshold [29, 30]. Similar saturation behavior has also been observed for microresonator frequency combs in the drop-port experiments of [5] and in simulation studies [6].

As a further test, we investigated a resonance belonging to the $TM_2$ family at ≈1551.5 nm with a relatively low Q (≈$6.5 \times 10^5$). For this resonance no comb was generated at up to 430 mW input power, the highest power tested. The pump power emerging from the drop-port was approximately linear with input power, with no sign of saturation. This supports our interpretation that the drop-port data of Fig. 3 provide evidence of parametric gain saturation directly connected to Kerr comb formation.

From Fig. 3 we also observe that the output power from the comb obtained by pumping at $TM_2$ is substantially higher than that for the comb obtained by pumping at $TM_1$. Furthermore, in both cases the output power in the comb is relatively small compared to the input power. We may understand these trends, at least in part, from the linear transmission spectra and coupling parameters. For the $TM_1$ resonance, the total drop-port output power at the highest pump level is 9.6 mW, quite close to the 12 mW value obtained from the product of the pump power (430 mW) and the linear drop-port transmission ($T_{drop}$ = 0.028). A similar estimate for the total drop-port power for the $TM_2$ resonance, using 680 mW pump power and $T_{drop}$ = 0.142, yields 97 mW, higher than but within a factor of two of the observed value (55 mW).

In addition to linear behavior, a saturation behavior is evident for the comb and total (comb plus pump) drop-port power data in both resonances, but most notably in Fig. 3(d). We attempt to explain this saturation, which to our knowledge has not been discussed in previous studies, through a simple, semi-empirical model. On resonance and assuming coupling parameters <<1, we first write

$$P_{thru} = [\sqrt{P_{bus}} - \kappa_e \sqrt{P_{uring}}]^2 \qquad (3a)$$

$$P_{drop} \approx \beta(P_{bus} - P_{thru}) \qquad (3b)$$

where $P_{uring}$ is the resonantly enhanced power in the microring at the pump frequency. Eq. (3a) accounts for the interference between the field transmitted directly to the thru-port and the field coupled out of the microring and applies with or without comb generation. Eq. (3b) says that the drop-port power is equal to some coefficients $\beta$ multiplied by the power $P_{bus}$-$P_{thru}$ that does not go to the thru-port. For light that remains at the pump frequency, we would have $\beta_{pump} = \kappa_d^2 / (\kappa_p^2 + \kappa_d^2)$, whereas for light generated in the comb we would have $\beta_{comb} = \kappa_d^2 / (\kappa_p^2 + \kappa_d^2 + \kappa_e^2)$. The difference arises because the pump light delivered to the thru-port is already accounted for in Eq. (3a), while the comb light is not. Thus, as a larger fraction of the pump light in the microring is transferred into the comb, the effective β factor decreases. For the $TM_2$ resonance, this mechanism could eventually account for a ≈20% droop in the drop-port power. A second mechanism is related to the observed clamping of the pump light in the resonator. This alters the interference condition governing the transmitted pump at the thru-port; in Eq. (3a) we need to use the saturated value of the intracavity pump power rather than the linear value. Thus, increasing power transfer into the comb decreases the total power fraction coupled into our (under-coupled) resonators, presenting a second mechanism for saturation of the drop-port output power. These effects are significantly stronger for the $TM_2$ resonance, which is considerably closer to the critical coupling condition. Figure 4 shows the total drop-port power simulated for the $TM_2$ resonance according to Eq. (3). Here we use coupling parameters from Table 1 and take $\beta$ as the average of $\beta_{pump}$ and $\beta_{comb}$. We estimate the saturated pump power in the microring from the drop-port pump power data in Fig. 3(d). Because the pump power is not observed to saturate abruptly at threshold, we chose the value corresponding to a 280 mW input power for which the output pump power has clearly flattened out. Although the simulation result does not flatten out as much as the data in Fig. 3(d), it does exhibit a clear change in slope upon comb generation and predicts a total output power of 53 mW, quite close to the observed value (55 mW).

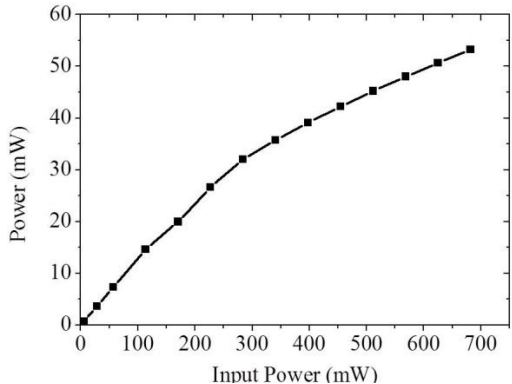

Fig. 4. The total drop-port power simulated for the TM$_2$ resonance according to Eq. (3).

We note that there are several factors that make a quantitative fit difficult. These include variation and uncertainty in the loss, especially the excess loss observed for the drop guide, the gradual (not abrupt) clamping of the intracavity pump power, and the variation of the $\kappa^2$ parameters (of order ±15% for the TM$_2$ mode) describing the resonances corresponding to the different lines within a single comb. We also cannot rule out additional nonlinear mechanisms. Nevertheless, we believe the simple model discussed explains the observed drop-port power data to a significant extent.

## 4. Time-domain characterization with drop-port geometry

The drop-port geometry also provides advantages for time-domain studies. Previous time-domain investigations characterized the comb at the thru-port [9, 10, 13-15]. In [9, 10, 13] line-by-line pulse shaping was used to compress the comb into a bandwidth-limited pulse train and demonstrate high coherence. Here the pulse shaper compensated the phase of the total field, including both the comb field as generated and the dispersion of the subsequent fiber link that relayed the comb to the measurement apparatus. However, because the dispersion of the fiber link was not characterized, no attempt was made to separate the initially generated comb waveform from subsequent dispersive reshaping. Subsequent investigations incorporated dispersion compensation into the fiber link and reported generation of ultrashort pulses directly in the microresonator [14, 15]. Note that [14, 15] both consider devices with anomalous group velocity dispersion. In all of these time-domain studies, suppression of the very strong pump line present at the thru-port, which would otherwise contribute a strong background field, was necessary. Reference [15] used a spectral filter to select a region of the comb away from the pump region for characterization, while in [9, 10, 13, 14] a pulse shaper or fiber-Bragg grating was used to attenuate the pump line. Here we upgrade our setup by dispersion compensating the fiber link and utilize the drop-port geometry to obtain a smooth spectrum directly from the microring. This enables time-domain characterization of the comb without the need for any intentional amplitude filtering. We report observation of ultrashort pulse generation directly from the microring, with low background and without filtering of the pump. Furthermore, to the best of our knowledge, our results are the first report of pulse generation directly at the microresonator for devices in the normal dispersion regime.

The autocorrelation trace of the comb generated from the TM$_1$ resonance, corresponding to the drop-port spectrum in Fig. 2(a), is plotted as the blue trace in Fig. 5. A pulse-like intensity autocorrelation is directly obtained. The measured pulse train has a period of 1.45 ps, corresponding to a 689 GHz repetition rate, consistent with the comb spacing of 5.52 nm (3 FSRs). The green trace in Fig. 5 shows the autocorrelation trace calculated on the basis of the optical spectrum after the EDFA and assuming flat spectral phase. The experimental and calculated traces are close but are not perfectly matched; experimental and simulated autocorrelations have widths of 526 fs and 423 fs, respectively (full-width half-maximum). One possible reason is that the pulses are short enough that residual higher-order dispersion in the fiber link causes some broadening. Another possibility is that for this case the field internal to the microring is itself slightly broadened compared to the bandwidth-limit.

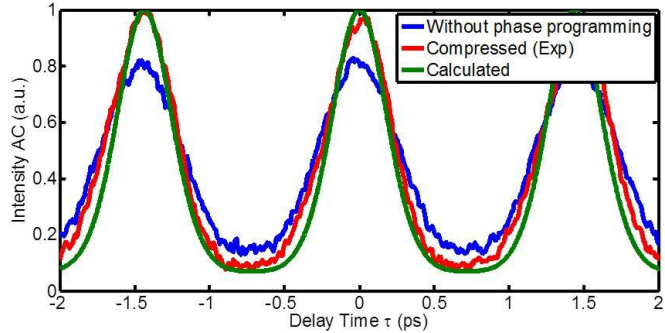

Fig. 5. Autocorrelation traces corresponding to the drop-port spectrum in Fig. 2(a). The green trace shows the intensity autocorrelation trace calculated by taking the optical spectrum and assuming flat spectral phase, while the blue and red traces show the measured autocorrelation traces before and after the line-by-line phase compensation.

For this particular comb, we also performed experiments in which a programmable pulse shaper is placed before the EDFA. The pulse shaper was used to fine tune the spectral phase, as in [9, 13], but without modifying the shape of the optical power spectrum. The measured autocorrelation trace after fine tuning the phase (red trace) is in close agreement with the calculation. This indicates a high level of coherence, in accord with previous studies of combs generated directly at multiple FSR spacing, e.g., [9].

We obtained a more dramatic example of a pulse-like intensity autocorrelation by pumping a different resonance of the same $TM_1$ mode family. Figure 6(a) and (b) show the optical spectra at 1 W pump power around 1560.4 nm, measured respectively (a) directly after the drop-port and (b) after the EDFA. The comb is formed directly with single free spectral range (FSR) line spacing, and the RF intensity noise of the generated comb (not shown here) is low, close to the background noise of the electrical spectrum analyzer (ESA). Again, a smooth spectrum is obtained, without the need for spectral filtering or pump attenuation after the microresonator. The measured autocorrelation trace, plotted as the blue line in Fig. 6(c), matches well with the trace simulated assuming flat spectral phase (green trace). The pulses have a period of 4.35 ps, corresponding to a 230 GHz repetition rate, consistent with the 1.84 nm comb spacing, with measured autocorrelation width ≈698 fs FWHM. Note that no pulse shaping is performed in this experiment. The close agreement apparent between experiment and simulation suggests passively mode-locked operation leading to nearly bandwidth-limited pulses internal to the microresonator.

It is worth reiterating that previous studies reporting passive mode-locking of the comb consistent with soliton-pulse formation were performed with devices in the anomalous dispersion regime [14, 15]. Most theories discussing conditions for comb formation are based on a modulational instability analysis, also assuming anomalous dispersion, e.g., [12]. A few recent studies provide possible explanations for comb generation in the normal dispersion regime, either due to a modification of the modulational instability process in the presence of a cavity with pump detuning or due to interactions with other mode families [31,

32]. Our results suggest that the mechanics of passively mode-locked pulse generation may also be operative for combs generated with normal dispersion.

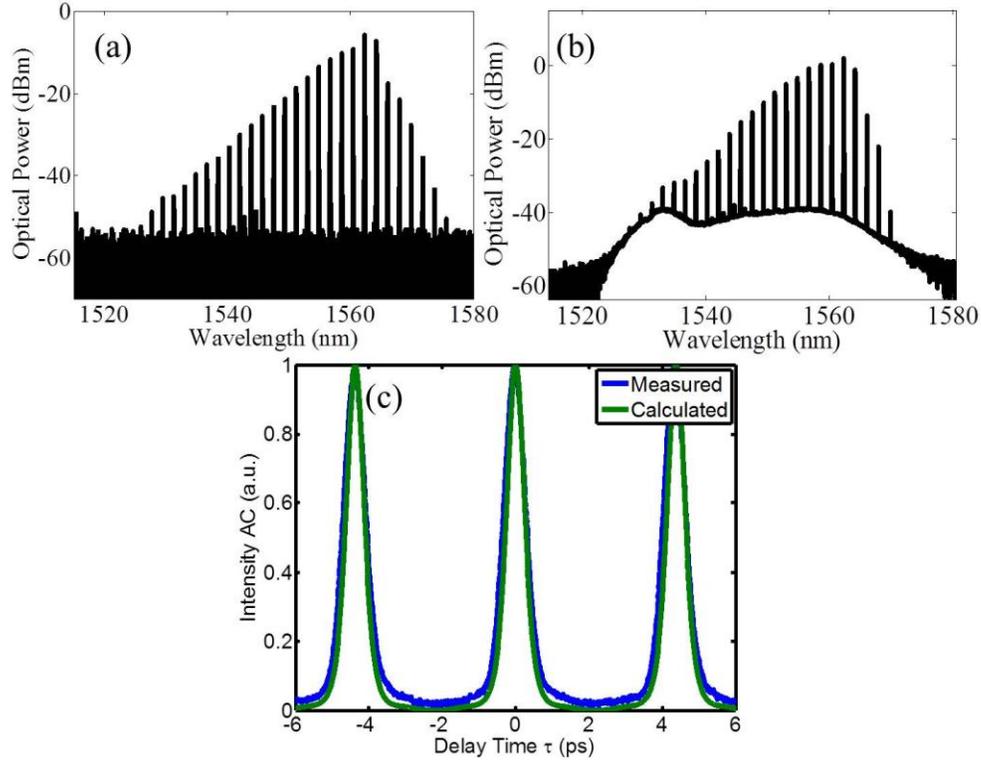

Fig. 6. (a) Optical spectrum measured from the drop-port with input power of 1 W at around 1560.4 nm. (b) The corresponding optical spectrum after the EDFA. (c) Measured (blue) and (c) calculated (green) autocorrelation traces.

## 5. Conclusion

We have demonstrated spectral pump equalization of on-chip microresonator frequency combs with a drop-port geometry. The ability to directly probe the optical spectra in the microresonator provides information on saturation both of the pump power and of the Kerr comb itself. By including the saturation of the pump power internal to the ring cavity in the coupling equations, we can offer an explanation for the nonlinear power dependence of the drop-port output. The insight gained is relevant for designing microresonators with coupling chosen to optimize power transfer to the combs. The smooth spectra obtained from the drop waveguide are also beneficial for the time-domain studies, allowing waveform characterization without the need to suppress the potentially large background associated with the strong residual pump in the thru-port geometry. Passively mode-locked pulses are observed in measurements taken at the drop-port of a normal dispersion microcavity.

**Acknowledgment**

This work was supported in part by the National Science Foundation under grant ECCS-1102110, by the Air Force Office of Scientific Research under grant FA9550-12-1-0236, and by the DARPA PULSE program through grant W31P40-13-1-0018 from AMRDEC.